\documentstyle[preprint,amstex,aps,eqsecnum,epsfig]{revtex}
\tighten

\def\a{\alpha}

\def\b{\beta}

\def\g{\gamma}

\def\L{\Lambda}

\def\l{\lambda}

\def\s{\sigma}
\def\om{\omega}
\def\Om{\Omega}
\def\t{\theta}

\def\bds#1{\boldsymbol{#1}}
\def\lowmp{\lower.11em\hbox{${\scriptstyle\mp}$}}

\def\vev#1{\langle #1 \rangle}

\def\lbare{\lambda_{\text{b}}}

\def\intL{\int_{-\Lambda}^\Lambda\frac{dk}{2\pi}}

\begin{document}
\preprint{ IFUM 660/FT-00 \,\, Bicocca-FT-00-10\bigskip}
\title{\bf The $O(\infty)$ nonlinear $\sigma$ model out of equilibrium} 

\author{{\bf C. Destri $^{(a,b)}$}  and {\bf E. Manfredini $^{(b)}$\bigskip}}

\bigskip

\address{
(a) Dipartimento di Fisica G. Occhialini, \\
   Universit\`a di Milano--Bicocca  and INFN, sezione di Milano$^{ 1,2}$
    \\
(b)  Dipartimento di Fisica,  Universit\`a di Milano \\ 
     and INFN, sezione di Milano$^{ 1,2}$
}
\footnotetext{
$^1$mail address: Dipartimento di Fisica, Via Celoria 16, 20133 Milano,
ITALIA.}
\footnotetext{
$^2$e-mail: claudio.destri@@mi.infn.it, emanuele.manfredini@@mi.infn.it
}
\date{November 2000}
\maketitle
\begin{abstract}
The out--of--equilibrium dynamics of the $O(N+1)$ nonlinear
$\sigma$--model in $1+1$ dimensions is investigated in the large $N$
limit. Regarding the nonlinearity as the effect of a suitable large
coupling limit of the $O(N+1)$ $\phi^4$ model, we first of all verify
that the two limits commute, so that the $O(\infty)$ nonlinear
$\sigma$ model is uniquely defined. Such model can be completely
renormalized also in the out--of--equilibrium context, allowing us to
study the consequences of its asymptotic freedom on the time evolution
far from equilibrium. In particular we numerically study the spectrum
of produced particles during the relaxation of an initial condensate
and find no evidence for parametric resonance, a result that is
consistent with the presence of the nonlinear contraint. Only a weak
nonlinear resonance at late times is observed.
\end{abstract}

\newpage
\section{Introduction and summary}
Recently, a great deal of attention has been paid to the relaxation of
a bosonic condensate in interaction with its quantum and/or thermal
fluctuations. Some of the main results of this research program have
been obtained in the description of the inflationary dynamics, where
one has to consider the expectation value of a scalar field rolling
down towards equilibrium. Since similar scalar theories can be used to
describe the low energy features of hadronic physics (like the
interaction among pions), this subject has been studied also in
connection with the formation of Disoriented Chiral Condensates (DCC),
which may happen after the collision of two ultrarelativistic nuclei.

Thus, much work has been done about the quantum evolution out of
equilibrium of the $\phi^4$ model in $3+1$ dimensions
\cite{bdh,devega2,relax,chkm,hkmp,chkmp}. As is well known \cite{wilson},
the renormalized theory is {\em trivial}. Practically, this means that
we should consider the model as an effective theory, keeping the
ultraviolet cut-off $\Lambda$ much smaller than some Landau scale. The
logarithmic dependence on $\Lambda$ should disappear from the renormalized
quantities, while a weak inverse power dependence remains. 

If we want to push the application of non equilibrium techniques to
more fundamental theories, like QCD, we should consider that the
ultraviolet properties change drastically. In those cases, in fact,
there is no Landau Pole in the ultraviolet and the renormalized
coupling becomes smaller and smaller as the momentum scale
increases. This corresponds to the property of asymptotic freedom,
whose presence justifies self--consistently the perturbative
renormalization procedure and allows in principle to perform the
infinite cut-off limit smoothly.

Motivated by this consideration and by its intrinsic relevance both in
Quantum Field Theory and in Statistical Mechanics, we analyse in this
paper the dynamical properties of the nonlinear $\sigma$ model in
$1+1$ dimensions, which is asymptotically free in the ultraviolet
\cite{zj}. Thus, $\Lambda$ can be pushed to infinity rigorously and
there should exist a renormalized out--of--equilibrium dynamics,
completely independent of the ultraviolet cut-off.

The linear and nonlinear $\sigma$ models in $3+1$ dimensions were
introduced in elementary particle theory in order to provide a useful
model of the low--energy strong interaction sector, which was able to
realize the $SU(2) \times SU(2)$ current algebra and the Partial
Conservation of Axial Current (PCAC) and satisfy the corresponding low
energy theorems \cite{gellmann}. Afterwards, the nonlinear $\sigma$
model has been considered fruitfully in many areas of Quantum Field
Theory and Statistical Mechanics, mainly in the description of quantum
spin chains and $2D$ spin models \cite{zj} and, quite recently, of
disordered conductors and of quantum chaos \cite{simons}.

We consider in this paper the $O(N)$ invariant model, in the large $N$
limit. First of all, we present in section \ref{eveq} an original
derivation of the relevant result that the large $N$ limit and the
large coupling limit, which turns the linear model in the nonlinear
one, commute. More precisely, we show that the same {\em classical}
(in the sense of Yaffe \cite{yaffe}) hamiltonian, which describes the
quantum dynamics of the model in the large $N$ approximation, is
obtained, no matter in which order we perform the two limits. Previous
studies on this subject were presented in \cite{chan}, using
perturbative techniques and the derivative expansion for the model in
3+1 dimensions, with the conclusion that the divergent terms are
universal, while finite parts do differ when taking the large coupling
limit on the quantum corrections on the linear model, or calculating
the same quantum corrections on the nonlinear model. In our case we
find instead that the large coupling limit of the $O(\infty)\,\phi^4$
model is completely equivalent to the large $N$ limit of the quantum
nonlinear $\sigma$ model. We also derive the evolution equations for
the nonlinear model at the leading order in the $1/N$ expansion, in
the case of an initial field condensate different from zero. We
implement the constraint by the use of a Lagrange multiplier, which we
denote $m^2$, since it enters the dynamics as a squared mass. We show
that the usual renormalization procedure, which makes the bare
coupling constant depend on the UV cutoff, is sufficient to get
properly renormalized, that is UV finite, evolution
equations. Moreover, we characterize the ground state of the model,
giving an interpretation of the dynamical generation of mass (the
so--called dimensional transmutation) in terms of a compromise between
energetic requirements and the constraint. We conclude by describing
suitable initial conditions for the condensate and the quantum
fluctuations. We want to emphasize that, while the approach we follow
in the study of the dynamical evolution in this field theory has
become by now quite standard and is very similar to that of ref.s
\cite{bdh,devega2,relax,chkm,hkmp,chkmp}, the different dynamical
properties of the nonlinear $\sigma$ model leads to different and new
results, most notably the absence of parametric resonances and the
consequent need for a detailed study of the scaling properties with
respect to variation of $\Lambda$, which we describe in the second
part of this work.

\begin{figure} 
\epsfig{file=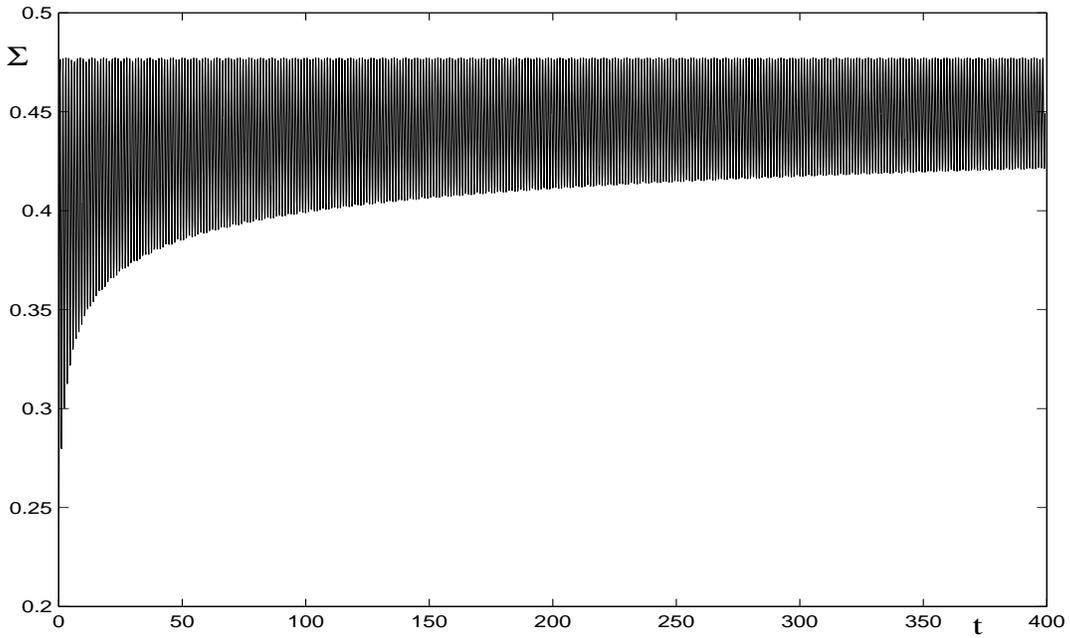,height=9cm,width=15cm}
\caption{\it Evolution of the backreaction $\Sigma (t)$ for $\L/m
_{\rm eq} =10$, $\ell=0$ and $\rho_0=0.5$.}\label{sigma-fig}
\end{figure}

In sec. \ref{num} we present the analysis of the numerical evolution
for the condensate and the Lagrange multiplier as well as for the
number of particles created during the relaxation of the condensate
(the quantum fluctuations). Remarkably, we do not find any period of
exponential growth for the fluctuations. Actually, no spinodal
instabilities were to be expected, since the symmetry is always
unbroken in 1+1 dimension.  But there occurs also no parametric
resonance, as takes place instead in the unbroken symmetry scenario of
the large$-N$ $\phi^4$ model in $3+1$ dimension. This is due to the
quite different nonlinearities of the $\s-$model and in particular to
the nonlinear constraint [see eq. (\ref{constr})] which sets an upper
bound to the quantum infrared fluctuations [see
fig. \ref{sigma-fig}]. In fact, even if the constraint disappears as
the bare coupling constant $\lbare$ vanishes in the infinite UV cutoff
limit (asymptotic freedom), the quantum fluctuations in any given
finite range of momentum remain constrained to finite values, as
implied by the possibility of fully renormalize the model, including
the constraint [see eq. (\ref{ren_lm})].  Because of this and of the
reduced momentum phase space, we observe that the damping of the
condensate is not as efficient as in the large$-N$ $\phi^4$ model in
$3+1$ dimension with unbroken symmetry. As a matter of fact our data
do not even allow to establish for sure that the condensate will eventually
relax to zero (in the case with zero angular momentum, see below).

In particular, we carefully study, by numerical fit and
self--consistent analytic computations, the asymptotic evolution of
the Lagrange multiplier. The estimated dependence of its asymptotic
value, $m(\infty)^2$, on the initial condensate $\rho_0$, turns out to
be very well approximated by an exponential, which is the exact
dependence of $m(0)^2$ [at infinite UV cutoff, see
eqs. (\ref{alfadiro}) and (\ref{m0dia})]; remarkably however, the
prefactor in the exponent is changed [see eq. (\ref{minfrho})]. We
also verify that, after the proper renormalization, the dependence of
the asymptotic values on the UV cutoff $\Lambda$ is only by inverse
powers. As far as the emission of particles is concerned, we
considered three different reference states: the initial state, the
adiabatic vacuum state and the equilibrium vacuum state, that is the
true ground state of the theory. The numerical results suggest a weak
nonlinear resonance, yielding a relaxation of the condensate via
particle production driven by power laws with non universal anomalous
exponents, a result similar to what found in \cite{relax} for the
asymptotic dynamics of $\phi^4$ in $3+1$ dimensions. However, more
numerical as well as analytical work is necessary for a better
quantitative estimates.

\begin{figure} 
\epsfig{file=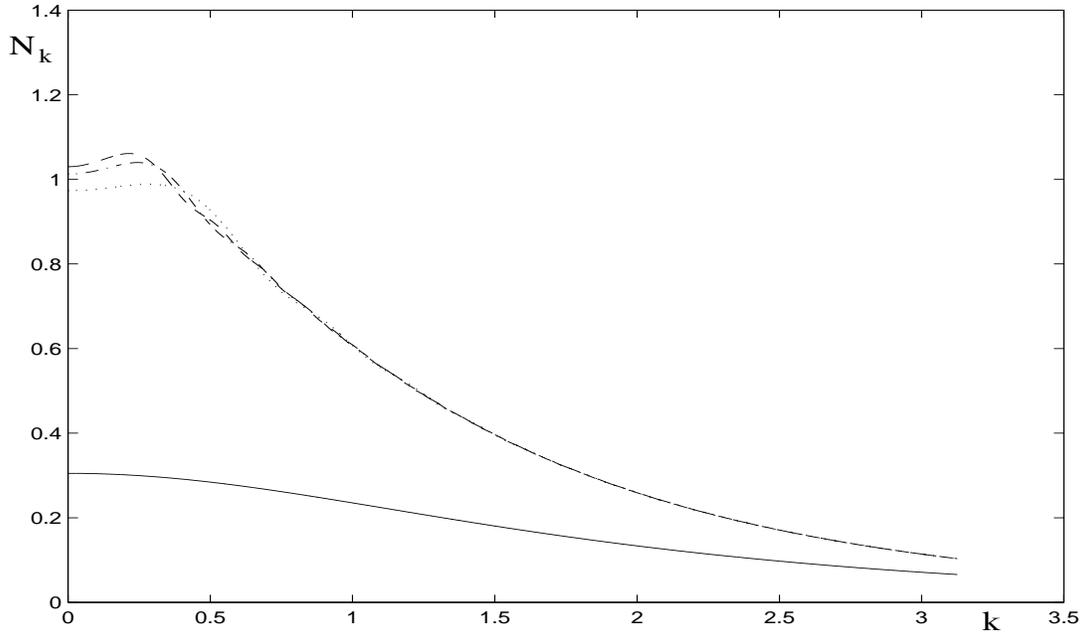,height=9cm,width=15cm}
\caption{\it Adiabatic spectrum for $tm_{\rm eq}=0.0$ (solid line),
$59.895$ (dotted line), $119.835$ (dashdot line) and $179.775$ (dashed
line), for $\L/m _{\rm eq} = 10$, $\ell=1$ and $\rho_0=0.3$.}
\label{lneq0_3}
\end{figure}

Finally, since we allow the condensate to have a number $n$ of components
larger than $1$, we are able to study the evolution of configurations
with non--zero angular momentum $\ell$ in the internal space of the
field [see eq. (\ref{eqm1})]. In this case we find numerical evidence
for an adiabatic spectrum broader than in the case $\ell=0$ [see
figure \ref{lneq0_3}], suggesting a stronger coupling with hard
modes. Again, more work is necessary for a better understanding of
this issue. 

There are, of course, physically relevant issues which are not
addressed in this paper, like the effects of subleading (in $1/N$)
terms, which may lead to genuine thermalization, and the consequent
non--uniformities between large time and large $N$, or the comparison
with the evolution of spatial inhomogeneous condensate \cite{in_homo}.

It nevertheless appears evident that, when compared to the
unconstrained $\phi^4$ model, the dynamics of this constrained model
has qualitatively different properties, which deserve a separated and
more detailed analysis.

\section{The $O(\infty)$ nonlinear $\sigma$ model in $1+1$ dimensions}
\label{eveq}

\subsection{Definitions}
\label{def}
The classical Lagrangian of the $O(N+1)$ $\s$ model is given by
\begin{equation}\label{clagr}
	L = \frac12 \, \partial_{\mu} \bds{\phi} \cdot 
	\partial_{\mu} \bds{\phi}
\end{equation}
where $\bds{\phi}$ is a multiplet transforming under the fundamental
representation of $O(N+1)$ and constrained to the $N-$dimensional
sphere of radius $\l^{-1/2}$:
\begin{equation}
	\phi^2 \equiv \bds{\phi} \cdot \bds{\phi} = 1/\l
\end{equation}
$\l$ may be regarded as the coupling constant, since the sphere
flattens out in the $\l\to0$ limit. The Hamiltonian corresponding to
(\ref{clagr}) reads
\begin{equation}\label{chami}
	H = \frac12 \int dx \left[ J^2 + (\partial_x\phi)^2 
	\right] \;,\quad J^2 = \sum_{i<j} J_{ij}^2
\end{equation}
where $J_{ij}=\phi_i\pi_j-\phi_j\pi_i$ is the angular momentum on the
sphere, $\pi_j$ being the momentum conjugated to $\phi_j$.  This
Hamiltonian can also be obtained as the $g\to\infty$ limit of the
linear model
\begin{equation}
	H_{\text{L}} = \int dx \left[ \frac12 \pi^2 + \frac12 
	(\partial_x\phi)^2 + V(\phi^2)\right]
\end{equation}
where $\bds\phi$ is now unconstrained and the potential $V$ may be taken
of the form
\begin{equation}
	V(u) = \frac{g}4 (u - 1/\l)^2
\end{equation}
The quantum version of the linear model defines a textbook
Quantum Field Theory (apart from the nontrivial strong coupling limit
$g\to\infty$). The quantum version of the nonlinear model
(\ref{chami}) may be written instead
\begin{equation}\label{qhami}
	\hat{H} = \frac12 \int dx \left[ -\bigtriangleup + 
	\,\om ^2\,(\partial_x \hat{\a})^2 \right]
\end{equation}
where we have used the projective coordinates $(\a_1,\ldots,\a_N)$
on the sphere, namely
\begin{equation}\label{proj}
	\phi_j = \om\,\a_j \;, \quad 
	\lbare^{1/2}\phi_{N+1} = \om - 1 \;,\quad
	\om = \frac{2}{1 + \lbare \a^2}
\end{equation}
so that
\begin{equation}
	(\partial_x\phi)^2 = \om ^2\,(\partial_x \a)^2 
\end{equation}
and the $O(N+1)-$symmetric functional Laplacian reads
\begin{equation}
	\bigtriangleup(x) = \om(x)^{-N} \frac{\delta}{\delta \a_j(x)}
	\,\om(x)^{N-2}\, \frac{\delta}{\delta \a_j(x)}
\end{equation}
while $\a$ in eq. (\ref{qhami}) is a multiplicative operator. We have
replaced the coupling constant $\l$ with $\lbare$ (the {\em bare}
coupling constant) to stress the fact that in Quantum Field Theory it
is generally cut-off dependent.

\subsection{The $N\to\infty$ limit}
\label{limit}
Now we derive the quantum dynamics in the large $N$ limit, applying a
general technique already used in the analysis of the $\phi^4$
dynamics in finite volume \cite{lN} and based on well--known work by
Yaffe \cite{yaffe}. If we consider the nonlinear model as a limit of
the $\phi^4$ linear model (being this true at least at the classical
level), we have to take two limits and we might wonder whether it is
legitimate to interchange their order. To be more specific, if we
first perform the large $N$ limit in the linear model, we get a
classical $g-$dependent unconstrained Hamiltonian $H^\infty_{\text{L}}$, that
admits a definite nonlinear limit $H^\infty$ as $g\to\infty$. We verify here
that indeed the same Hamiltonian $H^\infty$ follows if we start directly form
the nonlinear quantum Hamiltonian (\ref{qhami}) and take  
the $N\to\infty$ \`a la Yaffe. 

Consider the quantum Hamiltonian of the linear model, with the
couplings suitably rescaled to allow the large $N$ limit
\begin{equation}\label{lqhami}
	\hat{H}_{\rm L} = \int dx\left[ \frac12 \hat{\pi}^2 + 
	\frac12(\partial_x\hat{\phi})^2  + V(\hat{\phi}^2)\right] \;,\quad
	V(u) = \frac{g}{4N} (u - N/\lbare)^2 
\end{equation}
According to (a slight extension of) Yaffe's rules, the quantum
dynamics described by the $N \to \infty$ limit of the model is
described by a {\em classical} Hamiltonian, which is the large $N$
limit of the expectation value of the quantum hamiltonian
(\ref{lqhami}) on a set of generalize coherent states, labelled by the
parameters defined in eq. (\ref{Nlim}). We end up with the following
classical hamiltonian
\begin{equation}\label{phi_ham}
\begin{split}
	H^\infty_{\rm L} =& \lim_{N\to\infty}\frac{\vev{{\hat H}_{\rm L}}}N 
	= \int dx \left[ \frac12 \bds\pi ^2 + 
	\frac12(\partial_x\bds\phi)^2 + V(\bds\phi^2 + w(x,x))
	\right]\\ &+ \frac12 \int dx\,dx'\,dx''\,v(x,x')w(x',x'')
	v(x'',x)\\ &+ \int dx \left[ \frac18 w ^{-1} (x,x) - \frac12 
	\partial_x^2 w(x,x')\bigr|_{x'=x} \right] \;,\quad 
	V(u)=\frac{g}{4}(u - 1/\lbare)^2
\end{split}
\end{equation}
where the classical canonical variables are defined as
\begin{equation}\label{Nlim}
	\begin{pmatrix}
	   \bds\phi(x) \\ \bds\pi(x) \\ w(x,x') \\ v(x,x')
	\end{pmatrix} = \lim _{N\to\infty} \frac1{N}
	\begin{pmatrix}
	   \sqrt{N} \vev{\hat{\bds\phi}(x)} \\ 
	   \sqrt{N} \vev{\hat{\bds\pi}(x)} \\ 
	   \vev{\hat{\bds\phi}(x)\cdot \hat{\bds\phi}(x')}_{\rm conn}\\ 
	   \vev{\hat{\bds\pi}(x)\cdot \hat{\bds\pi}(x')}_{\rm conn}
	\end{pmatrix}
\end{equation}
and the nonvanishing Poisson brackets read
\begin{equation}\label{PB}
\begin{split}
	\left\{ \phi _j (x) \,,\pi _k (x') \right\}_{\rm P.B.} &=
	\delta _{jk} \delta( x - x' ) \\
	\left\{ w(x,y)\,, v(x',y') \right\}_{\rm P.B.} &=
	 \delta(x-x') \delta(y-y') + \delta(x-y') \delta(y-x')
\end{split}
\end{equation}
It is understood that the dimensionality of the vectors $\bds\phi(x)$
and $\bds \pi(x)$ is arbitrary but finite [that is, only a finite
number, say $n+1$, of pairs $(\hat{\bds\phi}(x)\,,\hat{\bds\pi}(x))$
may take a nonvanishing expectation value as $N\to\infty$]. Thus, the
index $j$ may run form $1$ to $n+1$, where $n+1$ is the number of
field components with non zero expectation value.

The $g\to\infty$ limit on the classical Hamiltonian $H^\infty_{\rm L}$ is
straightforward and reintroduces the spherical constraint in the new form
\begin{equation}\label{newcon}
	\sum _{j=1} ^{n+1} \phi _j ^2 + \text{diag}(w) = 1/\lbare
\end{equation}
whose conservation in time implies 
\begin{equation}\label{newconp}
	\bds \phi \cdot \bds\pi + \text{diag}(wv) = 0
\end{equation}
where we have introduced the condensed notation
\begin{equation}
	(ab)(x,y) \equiv \int dz\, a(x,z)\,b(z,y) \;,\quad
	\text{diag}(a)(x) \equiv a(x,x)
\end{equation}
Let us now come back to the quantum Hamiltonian (\ref{qhami}) of 
the non-linear model. First of all we perform a similitude
transformation of the Laplacian, to cast it in a form suitable for the
application of Yaffe's method:
\begin{equation}
	\!\!\!-\overline{\bigtriangleup} = -\om^{N/2} \bigtriangleup 
	\om ^{-N/2} = \frac{1}{2} \left(\om^{-2} {\hat \b}^2 + 
	{\hat \b}^2 \om^{-2} \right) + \left(\frac{1}{2} + 
	\frac{2}{N} \right) \frac{{\hat\a}^2}{N} - \frac{N}{4} + 1
\end{equation}
where $\hat\a_j(x)$ is the obvious multiplication operator and
$\hat\b _j(x)=-i\delta/\delta\a_j(x)$ its conjugated momentum. Now, after
the rescaling $\lbare\to\lbare/N$ in eqs. (\ref{proj}), by the
usual rules in the $N\to\infty$ limit we obtain the classical
Hamiltonian
\begin{equation}\label{sigma_ham}
\begin{split}
	H^\infty =& \frac12 \int dx \left\{\Om^{-2} \left[\bds\b^2 + 
	\text{diag}(\chi\eta\chi) + \frac14 \text{diag}(\eta^{-1}) 
	\right] \right.\\ &+ \left. \Om^2 \left[  (\partial_x\bds\a)^2
	+ \partial_x\partial_{x'}\eta(x,x')\bigr|_{x=x'}\right] \right\}
\end{split}
\end{equation}
where
\begin{equation}
	\Om = \frac{2}{1 + \lbare[\bds\a^2 + \text{diag}(\eta)]}
\end{equation}
and, just as in eq. (\ref{Nlim}),
\begin{equation}
	\begin{pmatrix}
	   \bds\a(x) \\ \bds\b(x) \\ \eta(x,x') \\ \chi(x,x')
	\end{pmatrix} = \lim _{N\to\infty} \frac1{N}
	\begin{pmatrix}
	   \sqrt{N} \vev{\hat{\bds\a}(x)} \\ 
	   \sqrt{N} \vev{\hat{\bds\b}(x)} \\ 
	   \vev{\hat{\bds\a}(x)\cdot \hat{\bds\a}(x')}_{\rm conn}\\ 
	   \vev{\hat{\bds\b}(x)\cdot \hat{\bds\b}(x')}_{\rm conn}
	\end{pmatrix}
\end{equation}
are classical canonically conjugated pairs, with Poisson brackets
identical to those in eq. (\ref{PB}). We take the indices of the
classical fields $\bds\a$ and $\bds\b$ to run form 1 to $n$, having
assumed that only the first $n$ components of their quantum
counterparts may have expectation values of order $\sqrt{N}$.

To show that the classical Hamiltonian $H^\infty$ is equivalent to the
$g\to\infty$ limit of $H^\infty_{\rm L}$, we need only to solve the spherical
constraint (\ref{newcon}) that emerges in that limit. This amounts to
the canonical parameterization of the constrained pairs
$(\bds\phi,\bds\pi)$ and $(w,v)$ in terms of the projective ones
$(\bds\a,\bds\b)$ and $(\eta,\chi)$. It reads
\begin{alignat}{2}
	\phi_j &= \lbare^{1/2}\Om\,\a_j \;,\qquad &
	\phi_{n+1} &= \Om - 1 \\
	\pi_j &= \Om^{-1} \b_j + \alpha _j\pi_{n+1} \;,\qquad &
	\pi_{n+1} &= - \bds\a\cdot\bds\b -\text{diag}(\eta\chi)\\
	w(x,x') &= \Om(x) \Om(x')\,\eta(x,x') \;,\quad &	
	v(x,x') &= \dfrac{\chi(x,x')}{\Om(x)\Om(x')} +
	\dfrac{\delta(x-x')}{\Om(x)} \pi_{n+1}(x)
\end{alignat}
and in particular it implies, besides (\ref{newcon}) and
(\ref{newconp}),
\begin{equation}
\begin{split}
	\bds\pi^2 + \text{diag}(vwv) &= \Om^{-2} \left[ 
	\b^2 + \text{diag}(\chi\eta\chi) \right] \\
	(\partial_x\bds\phi)^2 + \partial_x\partial_{x'}w(x,x')\bigr|_{x=x'} 
	&= \Om^2 \left[ (\partial_x\bds\a)^2 + 
	\partial_x\partial_{x'}\eta(x,x')\bigr|_{x=x'}\right]
\end{split}
\end{equation}
This result proves the complete equivalence between the $g\to\infty$
limit on the leading $1/N$ term of the linear model (which imposes the
new spherical constraint) and the $N\to\infty$ limit of the quantum
model directly formulated on the constraint manifold.

Before closing this section, it should be noticed that, even though we
gave the basic definitions and performed the entire computation for a
field theory in $1+1$ dimensions, the results in sections \ref{def} and
\ref{limit} remain valid also for a $(D+1)-$dimensional theory, the only
change being in the dimensionality of the integrals.

\subsection{Dynamical Evolution}
Let us now derive the evolution equation for this system in the case
the field $\hat{\bds\phi}$ has a non zero, albeit uniform, expectation
value $\bds\phi$ in the initial state. The $2-$point
functions depend only on the difference $x-x'$, and can be
parametrized by time--dependent widths $\s_k$:
\begin{equation}\label{trinv}
	w(x,x') = \intL \, \s_k^2 \,e^{ i  k(x-x')}
	\;,\quad  v(x,x') = \intL \, 
	\frac{\dot\s_k}{\s_k} \,e^{ i  k(x-x')}
\end{equation}
where $\Lambda$ is the ultraviolet cut-off. 

In this case of translation invariance, in practice one can always
take $n=1$ owing to the $O(n+1)$ symmetry of $H^\infty_{\rm L}$. Thus,
we choose $\bds\phi$ to have only two non--zero components. In other
words the condensate will move on the plane specified by the initial
conditions for $\bds\phi$ and its velocity. Using eq. (\ref{trinv}),
we may write the Lagrangian density corresponding to the ($g\to\infty$
limit) of the Hamiltonian (\ref{phi_ham}) as
\begin{equation}
\begin{split}
	L = \frac{1}{2} \left( \dot{\phi} _{1} ^{2} +\dot{\phi}_2^2
	 \right) +& \frac{1}{2} \intL\left(\dot{\s}_k^2 - k^2 \s_k^2 - 
	\frac1{4 \s_k^2} \right)\\ - &\frac{m^2}{2} 
	\left(\phi_1^2 + \phi_2^2 + \intL \s_k^2 -\frac1{\lbare} \right)
\end{split}
\end{equation}
We have kept into account the constraint by introducing the Lagrange
multiplier $m^2$. The corresponding Euler--Lagrange evolution
equations read, in polar coordinates
\begin{align}
	& \ddot{\rho} + m ^2 \rho - \frac{\ell^2}{\rho ^3} = 0
	\label{eqm1}\\
	& \ddot{\s}_k + \left( k ^2 + m ^2 \right) \s _k - 
	  \frac{1}{4 \s _k ^3} = 0 \\ 
	& \rho ^2 + \Sigma - \frac{1}{\lbare} = 0 \label{constr}
\end{align}
with the definitions $\ell = \rho ^2 \dot{\theta}$ (the conserved angular
momentum of the condensate) and 
\begin{equation}\label{sigma}
	\Sigma = \text{diag}(w) = \intL \s_k^2
\end{equation}
The first thing we can do is to look for the minimum of the
Hamiltonian, that is the ground state of the theory which corresponds to the
vanishing $\ddot{\rho}$, $\dot{\rho}$, $\ddot{\s} _k$, $\dot{\s}_k$
and $\ell$. The equations to solve are:
\begin{equation}
\begin{split}
	& m ^2 \rho = 0 \\
	( &k ^2 + m ^2)\,\s_k - \frac{1}{4 \s _k ^3} = 0
\end{split}
\end{equation}
The solution $m = 0$ is not acceptable, because it yields a massless
spectrum for the fluctuations and gives an infrared divergence that
violates the constraint. This is nothing else than a different
formulation of the well--known Mermin-Wagner-Coleman theorem stating
the impossibility of the spontaneous symmetry breaking in $1+1$
dimensions \cite{mwc}. Thus, the unique solution is: $\rho = 0$ and
$\s _k = \frac12 (k ^2 + m ^2)^{-1/2}$.

This result allows for an interpretation of the mechanism of dynamical
generation of mass as the competition between the energy and the
constraint: in order to minimize the ``Heisenberg'' term in the
Hamiltonian, the zero mode width, that is $\s _0$, should be as
large as possible; on the other hand, it cannot be greater than a
certain value, because it must also satisfy the constraint. The
compromise generates a mass term, the same for all modes, which we
call $m_{\rm eq}$

We can take the mass at equilibrium as an independent mass scale
defining the theory, as dictated by the dimensional transmutation, and
the relation between this mass scale and the bare coupling constant is
read directly from the constraint (\ref{constr}) 
\begin{equation}\label{lbare}
	\frac{1}{\lbare} = \frac{1}{2\pi} \log\left(\frac{\Lambda}
	{m_{\rm eq}} + \sqrt{1 + \frac{\Lambda ^2}{m_{\rm eq}^2}} \right) =
	\frac{1}{2 \pi} \log \frac{2 \Lambda}{m_{\rm eq}} + O\left(
	\frac{m_{\rm eq}^2}{\Lambda ^2} \right)
\end{equation}

When the system is out of equilibrium, the Lagrange multiplier $m$ may
depend on time. Its behavior is determined by the fact that the
dynamical variables must satisfy the constraint. After some algebra,
this parameter can be written as:
\begin{equation}\label{Lagrange}
	m^2 = \lbare \left( \dot{\rho}^2 + \frac{\ell^2}{\rho^2} + 
	\Theta 	\right) \;,\quad \Theta = \intL
	\left(\dot{\s}_k^2 - k^2 \s_k^2 + \frac{1}{4\s_k^2} \right)
\end{equation}
We can describe the quantum fluctuations also by complex mode functions $z _k$,
which are related to the real function $\s _k$ by:
\begin{equation}\label{zdot}
	z_k  = \s _k e^{ i  \t_k} \;,\quad 
	\s_k^2\,{\dot\t}_k = \ell_k=\frac12 \;,\quad  
	\left| \dot{z}_k \right| ^2 = \dot{\s}_k ^2 + 
	\frac{\ell_k^2}{\s_k^2} 
\end{equation}
One can recognize in the second term on the r.h.s. of
the last equation in (\ref{zdot}) the centrifugal energy induced by
Heisenberg uncertainty principle. 

We choose the following initial conditions for this complex mode
functions:
\begin{equation}\label{init_fl}
	z_k(0) = \frac{1}{\sqrt{2 \om _k}} \;,\quad
	{\dot z}_k(0) = - i \sqrt{\om_k/2}
\end{equation}
where $\om_k =\sqrt{k ^2 + \alpha ^2}$ and $\alpha$ is an initial
mass scale. It is worth noticing here
that such a form for the initial spectrum of the quantum fluctuations
does not allow for an initial radial speed for the condensate degrees
of freedom, unless we start from $\rho_0=0$. This is easily seen by
differentiating (\ref{constr}) with respect to time. 

Moreover, we should stress that $\a$ might be different from the
initial value of the Lagrange multiplier. In fact, once the initial
value for $\rho$ is fixed, $\alpha$ can be determined by means of the
constraint equation and it turns out to be
\begin{equation}\label{alfadiro}
\begin{split}
	\alpha(\rho_0) &= m_{\rm eq} \exp \left(2 \pi \rho_0^2
	\right) \left\{\frac12 \left[ 1+ \sqrt{1+\frac{m_{\rm eq}^2}{\L^2}} +
	\exp (4 \pi \rho_0^2) \left( 1- \sqrt{1+\frac{m_{\rm
	eq}^2}{\L^2}} \right) \right] \right\}^{-1} \\ &= m_{\rm eq} \exp
	\left(2 \pi \rho_0^2 \right) \left[ 1 + O\left( \frac{m_{\rm
	eq}^2}{\Lambda ^2} \right) \right] 
\end{split}
\end{equation}
On the other hand, the initial value for the Lagrange multiplier is
given by
\begin{equation}\label{m0dia}
m^2_0 = \alpha^2 + \lbare \left( \stackrel{\cdot}{\rho} ^2 _0 +
\frac{l ^2}{\rho ^2 _0} - \a^2 \rho_0^2 \right)
\end{equation}
that is equal to the initial mass scale $\alpha ^2$ only if we push
the ultraviolet cut--off to infinity.

To properly control for any time the ultraviolet behavior of the
integrals in eqs. (\ref{sigma}) and (\ref{Lagrange}), one should
perform a WKB analysis \cite{bdv} of the solution. One finds the
following asymptotics for the mode functions:
\begin{equation}
	z_k(t) = z_k(0) \exp \left\{ikt - 
	\frac{ i }{2 k} \int_0^t dt'\, m^2(t') - 
	\frac{1}{4k^2} \left[ m^2(t) - m^2(0) \right] \right\} 
	\left[ 1 + O \left( \frac{1}{k^3} \right) \right]
\end{equation}
From the above formula it is clear that the logarithmic ultraviolet
divergence in $\Sigma$ is completely determined by the initial
spectrum. For the divergent integral $\Theta$ in eq. (\ref{Lagrange}) the
situation is more involved. Explicitly one finds:
\begin{equation}
	\Sigma(t) \equiv \intL |z_k(t)|^2 = 
	\frac{1}{2 \pi} \log \frac{\Lambda}{\mu} + \Sigma_{\rm F}(\mu;t) 
\end{equation}
and
\begin{equation}
	\Theta(t)=\intL \left(|{\dot z}_k(t)| ^2 - k^2 
	|z_k(t)|^2 \right) = m(t)^2 \,\Sigma(0) + \Theta_{\rm F}(t)
\end{equation}
where 
\begin{equation}
\begin{split}
	\Sigma_{\rm F}(\mu;t) &= \intL \left[ |z_k(t)|^2 - 
	\frac{\t(|k| - \mu)}{2|k|} \right] \\
	\Theta_{\rm F}(t) &= \intL\left[ |{\dot z}_k(t)|^2 - 
	k^2 |z_k(t)|^2-m(t)^2|z_k(0)|^2 \right]
\end{split}
\end{equation}
have finite limits as $\Lambda\to\infty$. We have introduced in the
above formulae a subtraction point $\mu$. There correspond a
renormalized coupling constant $\l$ running with $\mu$, as the
$\Lambda\to\infty$ limit of the relation
\begin{equation}
	\frac{1}{2 \pi} \log \frac{\Lambda}{\mu} - 
	\frac{1}{\lbare(\Lambda)} + \frac{1}{\l(\mu)} = 0
\end{equation}
and a renormalized constraint
\begin{equation}
	\rho(t)^2 + \Sigma_{\rm F}(\mu;t) - \frac{1}{\l(\mu)} = 0
\end{equation}
With this definitions, the equilibrium mass scale $m _{\rm eq}$ 
can be written as
\begin{equation}
	m _{\rm eq} = 2 \mu \exp \left[-\frac{2\pi}{\l(\mu)}\right]
\end{equation}
which by consistency with eq. (\ref{alfadiro}) implies
\begin{equation}
	\l(\a/2) = 1/\rho_0^2
\end{equation}
In conclusion we can rewrite the constraint and the Lagrange
multiplier as
\begin{equation}\label{ren_lm}
	\rho^2 + \Sigma_{\rm F}(\mu) - \frac{1}{\l(\mu)} = 0 \;,\quad
	m^2 = \frac{1}{\rho_0^2} \left[{\dot\rho}^2 + 
	\frac{\ell^2}{\rho^2} + \Theta_{\rm F}\right]
\end{equation}
For large but finite UV cutoff these expressions retain a inverse
power corrections in $\Lambda$. In the actual numerical calculations 
whose results will be presented in the following section, we used the
``bare'' counterparts of eqs (\ref{ren_lm}) with finite cutoff
and the definition (\ref{lbare}) of the bare coupling constant is used
to reduce to inverse power the cutoff dependence.

Let us conclude this section by summarizing the steps we need to do,
before trying to solve numerically the equations of motion. Once we
have fixed the UV cutoff, the equilibrium mass scale $m_{\rm eq}$ and
the initial value for the condensate $\rho_0$, we can determine the
initial mass scale $\a$ in the fluctuation spectrum from
eq. (\ref{alfadiro}), which in turn gives the initial conditions for
the complex mode functions [cfr. eq. (\ref{init_fl})]. Now, we need to
specify the remaining initial values for the condensate, namely its
velocity ${\dot\rho}_0$ and its angular momentum $\ell$, which must be
consistent with the constraint (\ref{newconp}). Finally,
eq. (\ref{ren_lm}) completely determines the initial value for the
Lagrange multiplier $m_0$, which has exactly the same infinite cutoff
limit as $\a$, but differs significantly from it for finite cutoffs.

\section{Numerical results}
\label{num}
We have studied numerically the following evolution equations
\begin{equation}
\begin{split}
	&{\ddot\phi} + m^2 \phi= 0 \\
	&{\ddot z}_k + \left( k ^2 + m ^2 \right) z _k = 0 \\
	&\frac{m^2}\lbare = |\dot\phi|^2  + \intL \left(|{\dot z}_k|^2
	 - k ^2 |z_k| ^2 \right)
\end{split}
\end{equation}
where $\phi = \phi_1+i\phi_2 = \rho\,e^{i\t}$, $\rho^2\,\dot\t = \ell$
and $|\dot\phi|^2 = \dot{\rho}^2 + \ell^2/\rho^2$, while the bare
coupling constant $\lbare$ is given by eq. (\ref{lbare}). The initial
conditions for $\phi$, $\dot\phi$ and $z_k$ [see eq.s (\ref{init_fl})]
must satisfy the constraints (\ref{newcon}) and (\ref{newconp}), that
are then preserved by dynamics.
 
In the classical limit the quantum fluctuations $z_k$ disappear from the
dynamics. In that case the stationary solutions are trivial: 
\begin{equation}
	\rho (t) = \lbare ^{-1/2} \;,\quad  m(t) = \lbare \ell
\end{equation}
with arbitrary value for the angular momentum $\ell$. Thus there are
stationary solutions corresponding to circular motion with constant
angular velocity.

When we include the coupling with quantum fluctuations, we still
obtain stationary solutions, parametrized by $\ell$ which assumes
arbitrary positive values. They have the following form:
\begin{equation}\label{stpt}
	\rho (t) =\sqrt{\frac{\ell}{m _{\rm eq} x}} \; \quad \; m (t) = m _{\rm eq} x
\end{equation}
where $x$ depends on $\ell$ through 
\begin{equation}
	\frac{2\pi\ell}{m_{\rm eq}x} +\sinh^{-1}\left(\frac\L{m_{\rm eq}x}
	\right) = \sinh^{-1}\left(\frac\L{m_{\rm eq}} \right) 
\end{equation}
which reduces to $x\log x=2\pi\ell/ m_{\rm eq}$ in the infinite
cut--off limit.

\subsection{Evolution of condensate and Lagrange multiplier}
\begin{figure} 
\epsfig{file=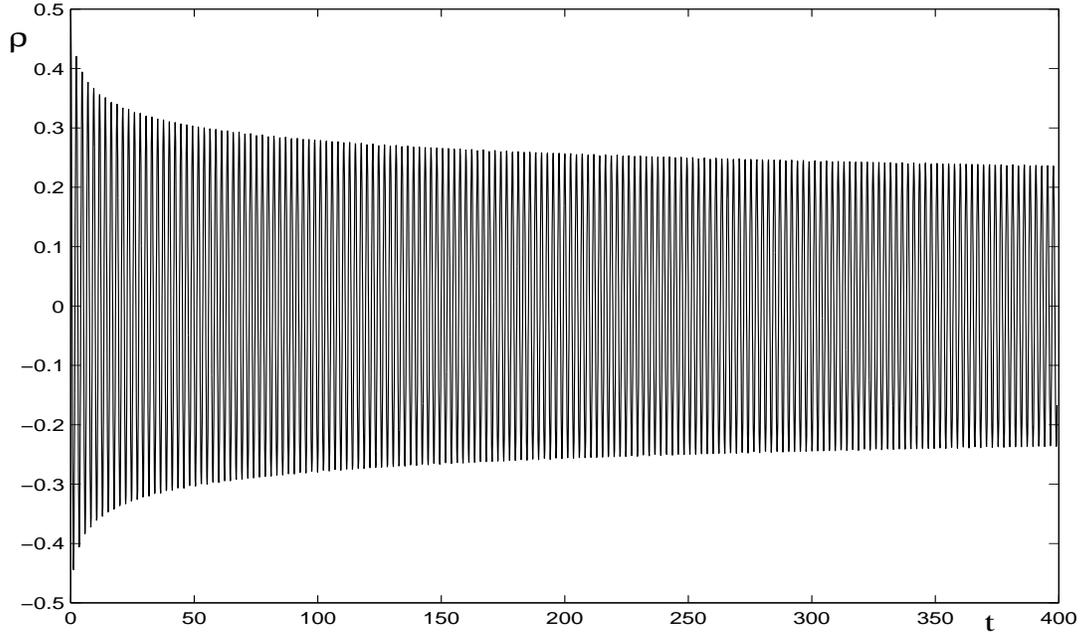,height=9cm,width=15cm}
\caption{\it Evolution of the mean value $\rho(t)$ for $\L/m
_{\rm eq} =10$, $\ell=0$ and $\rho_0=0.5$.}\label{ro}
\end{figure}

\begin{figure} 
\epsfig{file=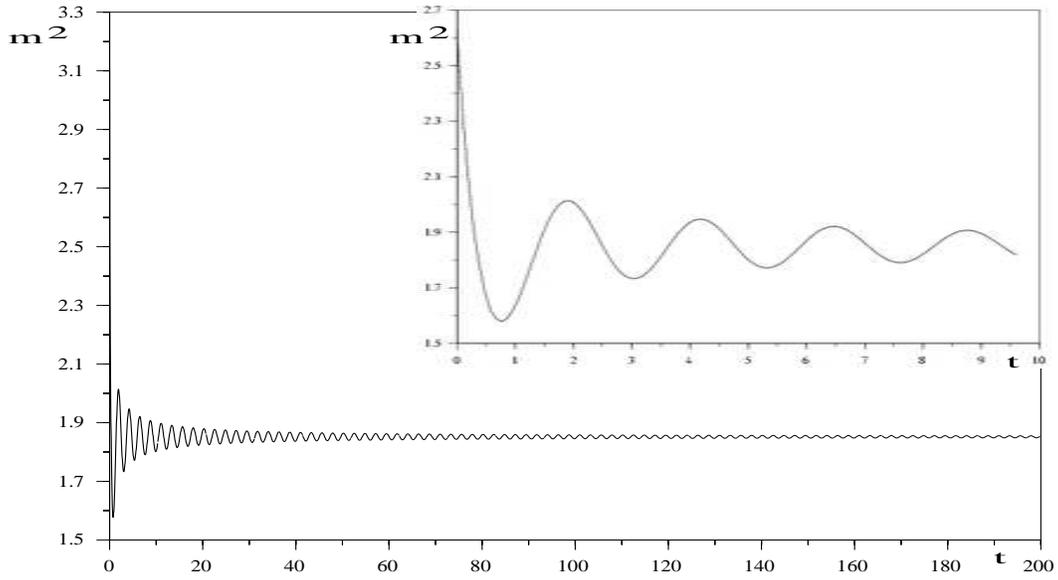,height=9.5cm,width=16.5cm}
\caption{\it Evolution of the Lagrange multiplier $m ^2 (t)$ for $\L/m
_{\rm eq} =20$, $\ell=0$ and $\rho_0=0.3$. In the smaller figure
there is zoom of the early times.}\label{lag_mult_ev}
\end{figure}

\begin{figure} 
\epsfig{file=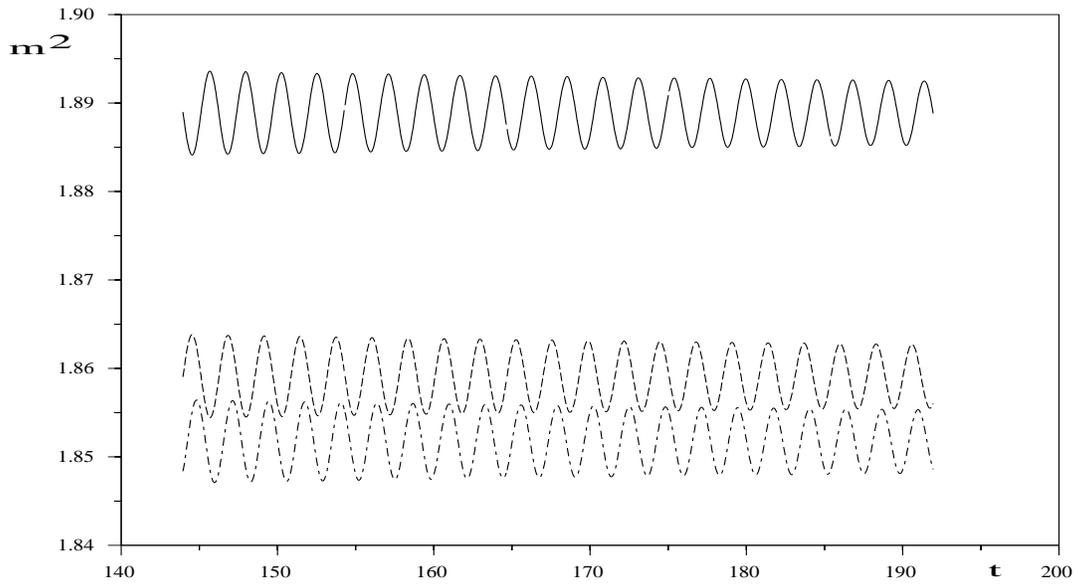,height=9.5cm,width=16.5cm}
\caption{\it Asymptotic evolution of $m ^2 (t)$ for three different
values of the ultraviolet cut--off: from top to bottom, $\L/m _{\rm
eq} = 5, 10$ and $20$, $\ell=0$ and $\rho_0=0.3$}\label{asy_sqmass}
\end{figure}

In order to control the dependence of the dynamics on the ultraviolet
cutoff, we solved the equations of motion for values of $\L$ ranging
from $5 m_{\rm eq}$ to $20 m _{\rm eq}$, with an initial condensate
ranging from $\rho_0 = 0.2$ to $\rho_0 = 0.7$. We mainly considered
the case $\ell=0$.  A typical example of the time evolution of the
relevant variables is showed in Fig.s \ref{ro} and
\ref{sigma-fig}. Figure \ref{lag_mult_ev} shows the evolution of the
Lagrange multiplier $m(t)^2$ for $\L/m _{\rm eq} = 20$; in this case,
its starting value is $2.630632$. Due to the lack of massless
particles, the damping of the oscillations of $\rho$ and $m^2$ is very
slow, as already noticed in \cite{chkm} for the linear model in $1+1$
dimensions; the dissipation is not as efficient as for the unbroken
symmetry scenario in $3+1$ dimensions, because of the reduced phase
space. A detailed numerical study of the asymptotic behavior and a FFT
analysis of the evolution allows a precise determination of the
asymptotic value and the main frequency of oscillation of the Lagrange
multiplier:
\begin{equation}\label{pert1}
m(t)^2 = m _{\infty} ^2 + \frac{p(t)}{t}  + O\left(\frac{1}{t^2}\right)
\label{asymass}
\end{equation}
where the function $p(t)$ turns out to be
\begin{equation}\label{pert2}
p(t) \simeq A \cos (2 m _{\infty} t + \gamma_1 \log t + \gamma_2) 
\end{equation}
The logarithmic dependence in the phase could be justified by
self--consistent requirements (see below), along the same lines of the
detailed calculations performed in ref. \cite{relax} in a similar
context. Numerically it is very difficult to extract and we do not
attempt it here. Comparing further our result with that reported in
ref. \cite{relax}, we should emphasize that we do not find any
oscillatory component of frequency $2 m_0$, as happens instead for the
$\phi^4$ model in $3+1$ dimensions. Moreover, as figure
\ref{asy_sqmass} shows, both the asymptotic mass $m _{\infty}$ and the
amplitude $A$ depend on the ultraviolet cutoff $\Lambda$. This
dependence may be fitted with great accuracy through a low order
polynomial in $1/\L^2$, showing that the standard renormalization
holds at any time, as anticipated by the WKB analysis. Therefore, the
extrapolated parameters $m_{\infty}^2$ and $A$ give us information on
the fully renormalized physical theory (in the large $N$
approximation). The table below collects the values of $m_{\infty}^2$
for different values of $\Lambda$ and of the initial condensate
$\rho_0$.  The last column contain the extrapolation to infinite
cutoff, obtained by the low order polynomial fit. The empty cells in
the last row correspond to a UV cutoff so small that the exact
$\alpha^2$ turns out to be negative; these values are excluded from
the fit.

\vskip .75truecm
\begin{tabular}{|c|c|c|c|c|c|c|c|c|c|}
\hline
$\rho _0$ & $\L=5$ & $\L=6$ & $\L=7$ & $\L=8$ &
$\L=9$ & $\L=10$ & $\L=11$ & $\L=12$ & $\L=13$ \\
\hline
$0.2$ & $1.3073$ & $1.3047$ & $1.3032$ & $1.3022$ & $1.3014$ &
$1.3010$ & $1.3006$ & $1.3004$ & $1.3001$ \\
\hline
$0.3$ & $1.8888$ & $1.8766$ & $1.8693$ & $1.8646$ &
$1.8614$ & $1.8591$  & $1.8574$ & $1.8561$ & $1.8551$ \\ 
\hline
$0.4$ & $3.3869$ & $3.3162$ & $3.2747$ & $3.2482$ &
$3.2303$ & $3.2175$ & $3.2082$ & $3.2011$ & $3.1956$ \\ 
\hline
$0.5$ & $8.7094$ & $7.9915$ & $7.6082$ & $7.3764$ & $7.2246$
& $7.1193$ & $7.0432$ & $6.9861$ & $6.9424$ \\ 
\hline
$0.6$ & $206.03$ & $52.433$ & $35.564$ & $29.2276$ & $25.969$ & $24.016$ 
& $22.732$ & $21.835$ & $21.178$ \\ 
\hline
$0.7$ & & & & & & & & & $238.12$ \\ 
\hline
\end{tabular}
\vskip 1cm
\begin{tabular}{|c|c|c|c|c|c|c|c|c|}
\hline
$\rho _0$ & $\L=14$ & $\L=15$ & $\L=16$ & $\L=17$ & $\L=18$ & 
$\L=19$ & $\L=20$ & $\L=\infty$ \\ 
\hline
$0.2$ & $1.300$ & $1.2998$ & $1.2997$ & $1.2996$ & $1.2995$ & $1.2995$
& $1.2994$ & $1.2989$\\
\hline
$0.3$ & $1.8543$ & $1.8536$ & $1.8531$ & $1.8527$ & $1.8523$ & 
$1.8520$ & $1.8517$ & $1.8493$ \\ 
\hline
$0.4$ & $3.1912$ & $3.1877$ & $3.1848$ & $3.1824$ & $3.1805$ & 
$3.1788$ & $3.1774$ & $3.1643$ \\ 
\hline
$0.5$ & $6.9080$ & $6.8804$ & $6.8580$ & $6.8395$ & $6.8241$ & 
$6.8111$ & $6.8001$ & $6.6990$ \\ 
\hline
$0.6$ & $20.682$ & $20.295$ & $19.988$ & $19.740$ & $19.536$ & 
$19.366$ & $19.223$ & $16.964$ \\ 
\hline
$0.7$ & $177.645$ & $146.523$ & $127.66$ & $115.07$ & $106.11$ &
$99.442$ & $94.302$ & $68.207$ \\
\hline
\end{tabular}
\vskip .75truecm

A similar table can be provided for the amplitude $A$ in the
eq. (\ref{pert2}). The values extrapolated to infinite cutoff in a
similar fashion as before, turn out to be:

\vskip .75truecm
\begin{tabular}{|c|c|c|c|c|c|c|}
\hline
$\rho_0$ & $0.2$ & $0.3$ & $0.4$ & $0.5$ & $0.6$ & $0.7$ \\
\hline
$A(\L=\infty)$ & $0.539$ & $0.701$ & $0.924$ & $1.39$ & $2.34$ &
$4.30$ \\
\hline 
\end{tabular}
\vskip .75truecm \noindent 
However, this fit is not as accurate as that for $m_{\infty}$.

It is interesting to observe that at large UV cutoff
$m_{\infty}$ has an exponential dependence on $\rho_0$ analogous to
that of $m_0$ (which coincides to $\a$ at $\Lambda=\infty$). Most
remarkably the prefactor in the exponent is modified by the time
evolution: we find 
\begin{equation}\label{minfrho}
	m_{\infty}^2 \sim \exp(2\g\, \rho_0^2) \;,\quad 
	3.5 \lesssim \g \lesssim 4.5
\end{equation}
The determination of $\g$ is rather rough due to the uncertainties
in the values of $m_{\infty}$ extrapolated to $\Lambda=\infty$ at larger
$\rho_0$. Notice in any case that the analog of $\g$ for $m_0$ is
$2\pi=6.28\ldots$. 

\begin{figure} 
\epsfig{file=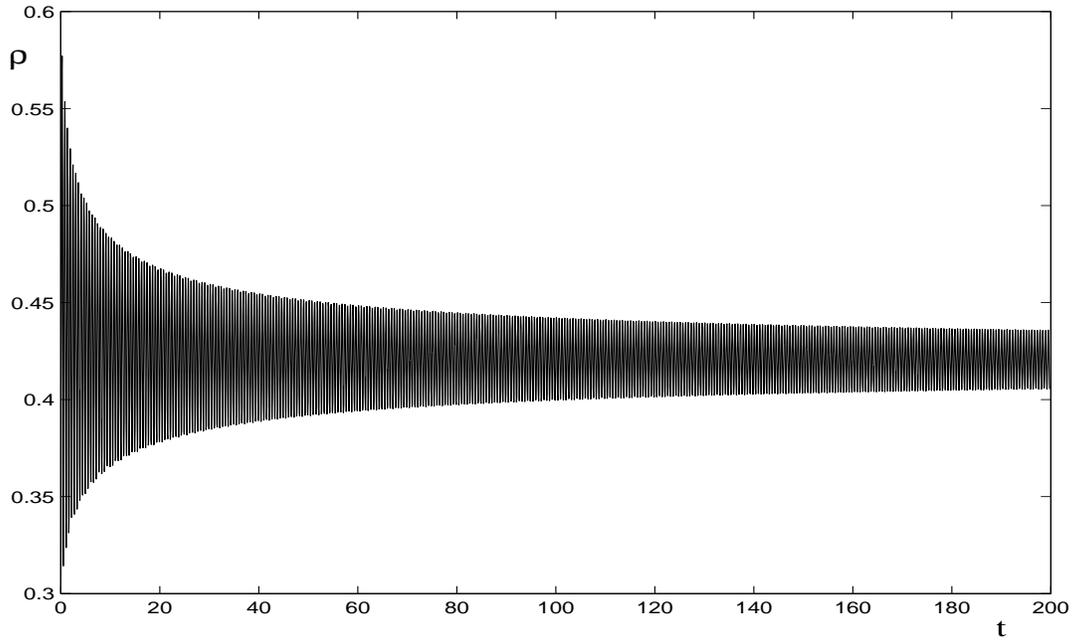,height=9cm,width=15cm}
\caption{\it Evolution of the mean value $\rho$ for $\L/m _{\rm eq} =
10$, $\rho_0=0.3$ and $\ell=1$.}
\label{lneq0_1}
\end{figure}

\begin{figure} 
\epsfig{file=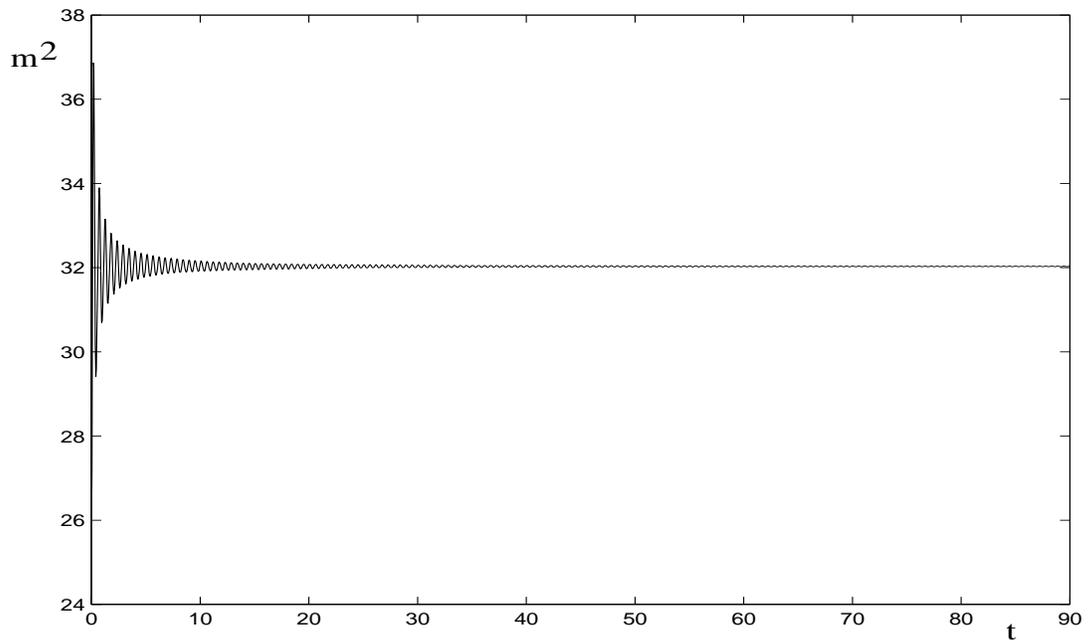,height=9cm,width=15cm}
\caption{\it Evolution of the squared mass $m^2$ for $\L/m _{\rm eq} =
10$, $\rho_0=0.3$ and $\ell=1$.}
\label{lneq0_2}
\end{figure}

We also performed some computations for $\ell>0$, with the following
results: if we start from an out of equilibrium value for $\rho$, it
will relax through emission of particles towards a fixed point,
different from the equilibrium value determined by eq.
(\ref{stpt}). Figures \ref{lneq0_1} and \ref{lneq0_2} show such a
situation for $\ell=1.0$, $\rho(0)=0.3$ and $\L = 10 m_{\rm eq}$. In that
case we have $x=1.000057$, while the mean values of the asymptotic
oscillations are $\rho _{\infty} = 0.4203$ and $m^2_{\infty} = 32.0294$.

\begin{figure} 
\epsfig{file=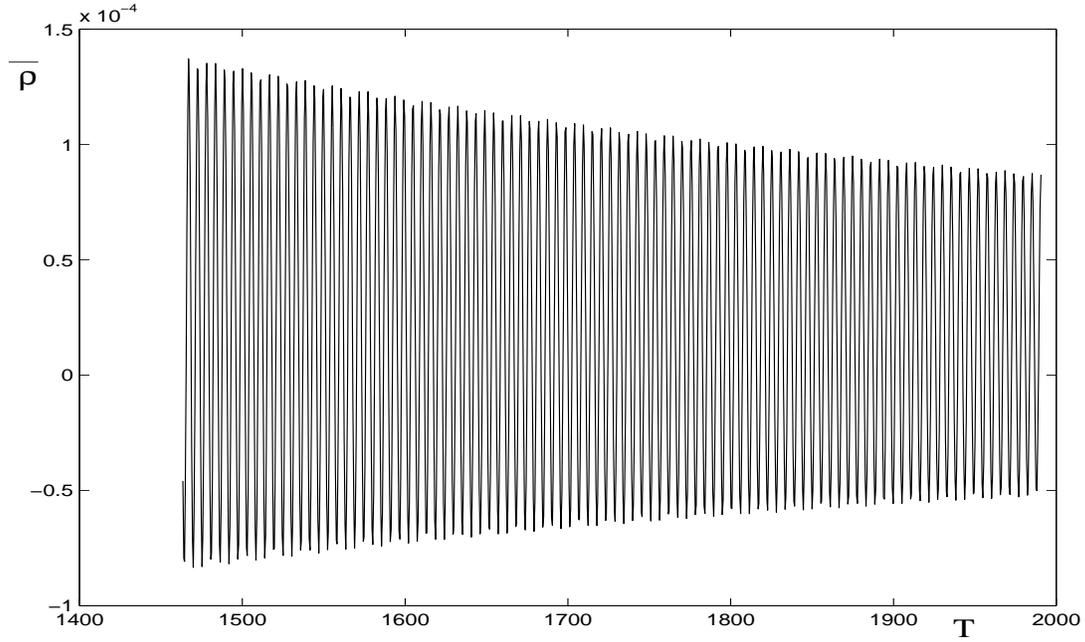,height=9cm,width=15cm}
\caption{\it The average value of the condensate $\rho$, defined as
$\bar{\rho}=\int^T\rho(t)dt/T$, plotted vs. $T$, for $\L/m_{\rm
eq}=20$, $\ell=0$ and $\rho_0=0.2$.}
\label{rob}
\end{figure}

\begin{figure} 
\epsfig{file=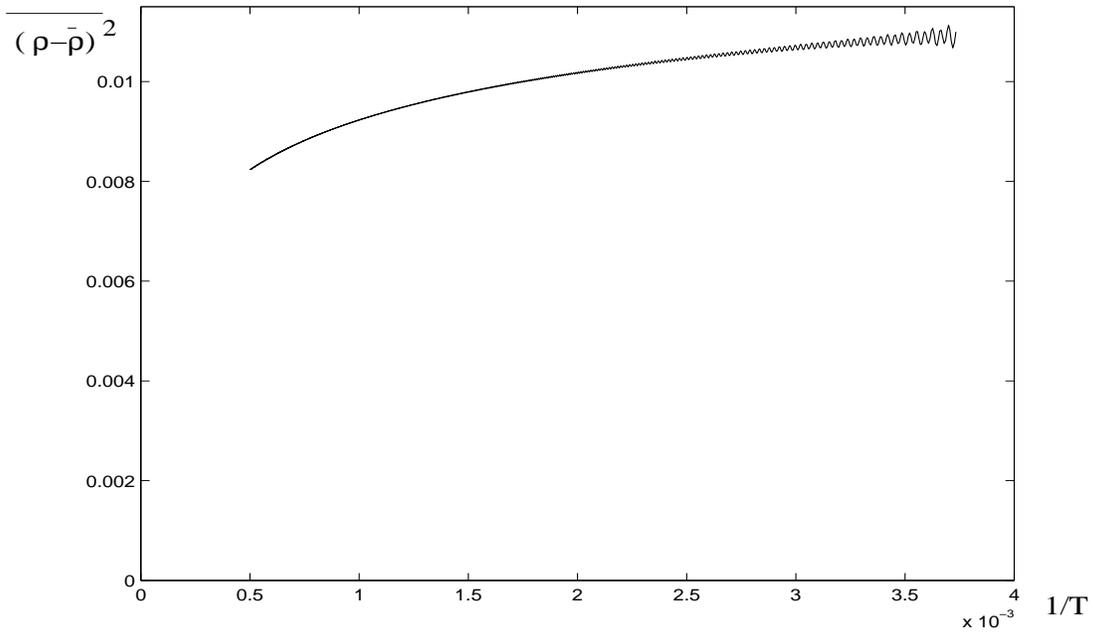,height=9cm,width=15cm}
\caption{\it The mean squared fluctuations of the condensate $\rho$, defined
as $\int^T(\rho(t)-\bar{\rho})^2dt/T$, plotted vs. $1/T$, for the same
values of the parameters as in figure \ref{rob}.}
\label{sq_flct}
\end{figure}

Before closing this section, we should comment a little further on the
evolution of the condensate $\rho$. When $\ell=0$, fig. \ref{rob}
shows that the oscillations are actually around zero. However, from
the available data, it is not possible to decide whether the amplitude
will eventually vanishes or will tend to a limiting cycle
(see fig. \ref{sq_flct}). 

On the other hand, in the case of $\ell
\neq 0$, it is already clear that the condensate does not relax to the
state of minimum energy compatible with the given value of $\ell$,
which would correspond to the circular orbit with radius given by
eq. (\ref{stpt}). However, it may still relax to a circular orbit with
a different radius and a different (larger) energy. More detailed and
longer numerical computations are needed to decide whether the damping
reduces the oscillation amplitude to zero or not.

\subsection{Emission spectrum}
\begin{figure} 
\epsfig{file=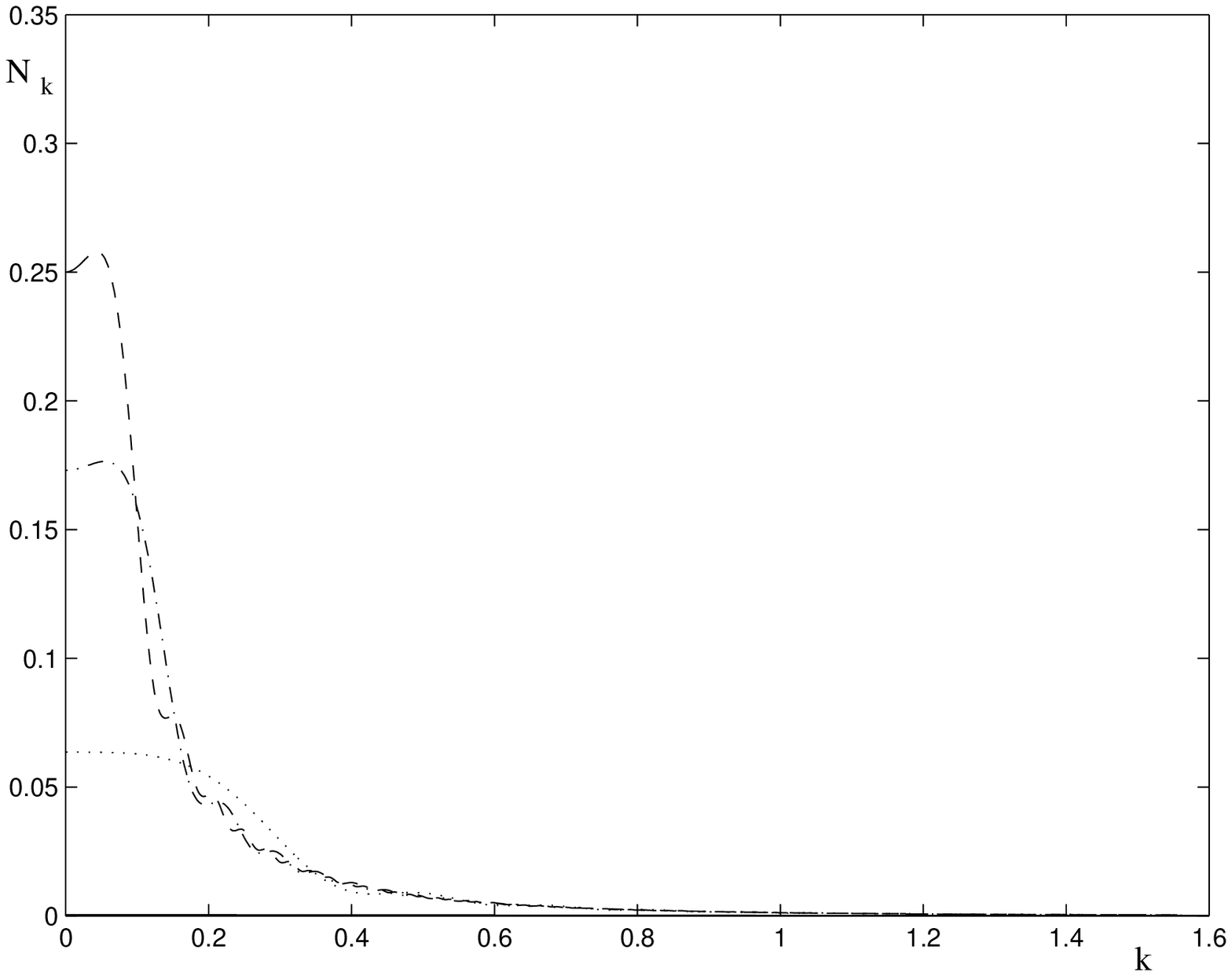,height=9cm,width=15cm}
\caption{\it Adiabatic spectrum for $tm_{\rm eq}=0$ (solid line),
$39.723$ (dotted line), $199.006$ (dashdot line) and $398.11$
(dashed line), for $\L/m _{\rm eq} = 20$, $\ell=0$ and $\rho_0=0.2$.}
\label{spec_1}
\end{figure}

\begin{figure} 
\epsfig{file=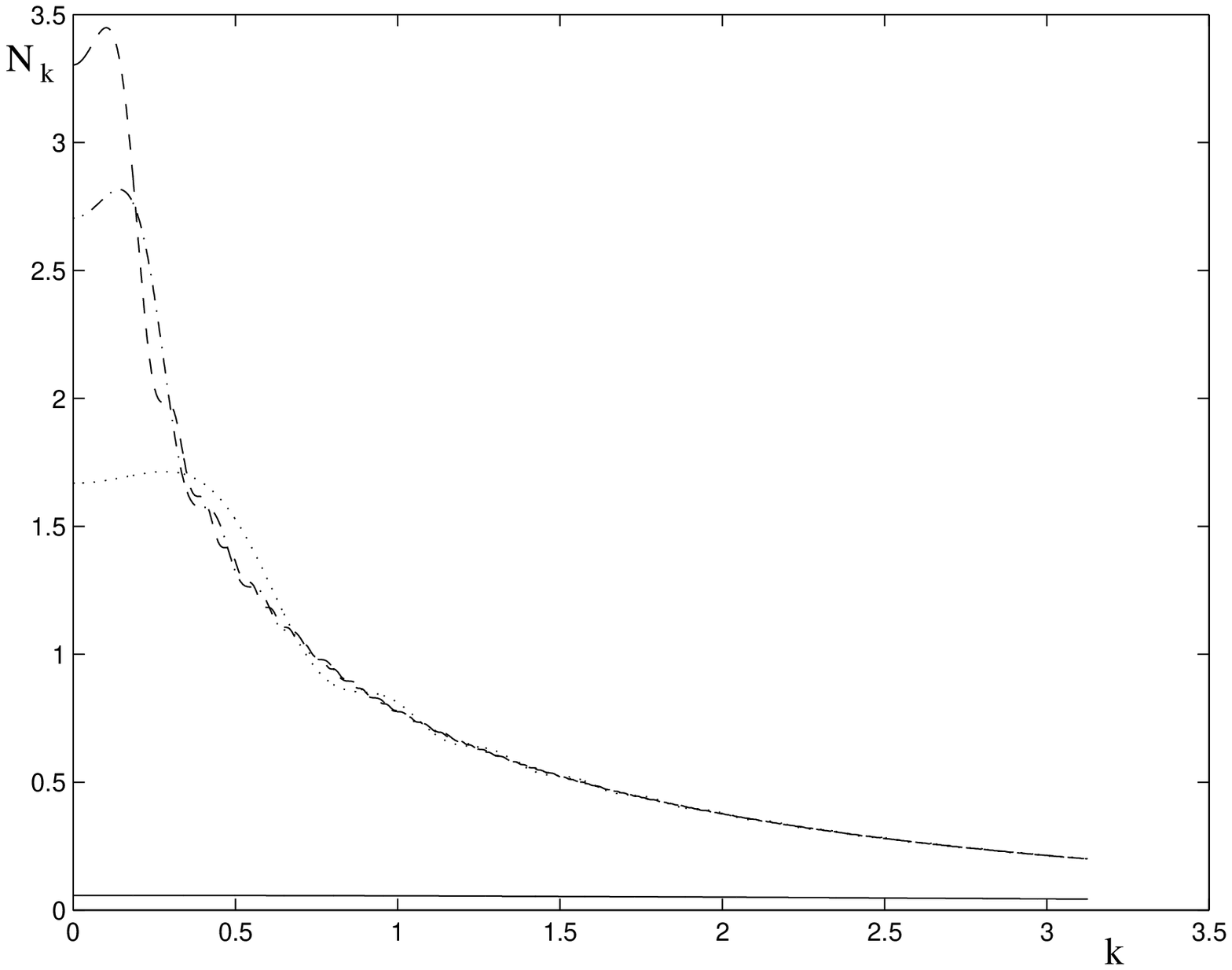,height=9cm,width=15cm}
\caption{\it Adiabatic spectrum for $tm_{\rm eq}=0$ (solid line),
$39.723$ (dotted line), $199.006$ (dashdot line) and $398.11$
(dashed line), for $\L/m _{\rm eq} = 20$, $\ell=0$ and $\rho_0=0.6$.}
\label{adnum0_6}
\end{figure}

\begin{figure} 
\epsfig{file=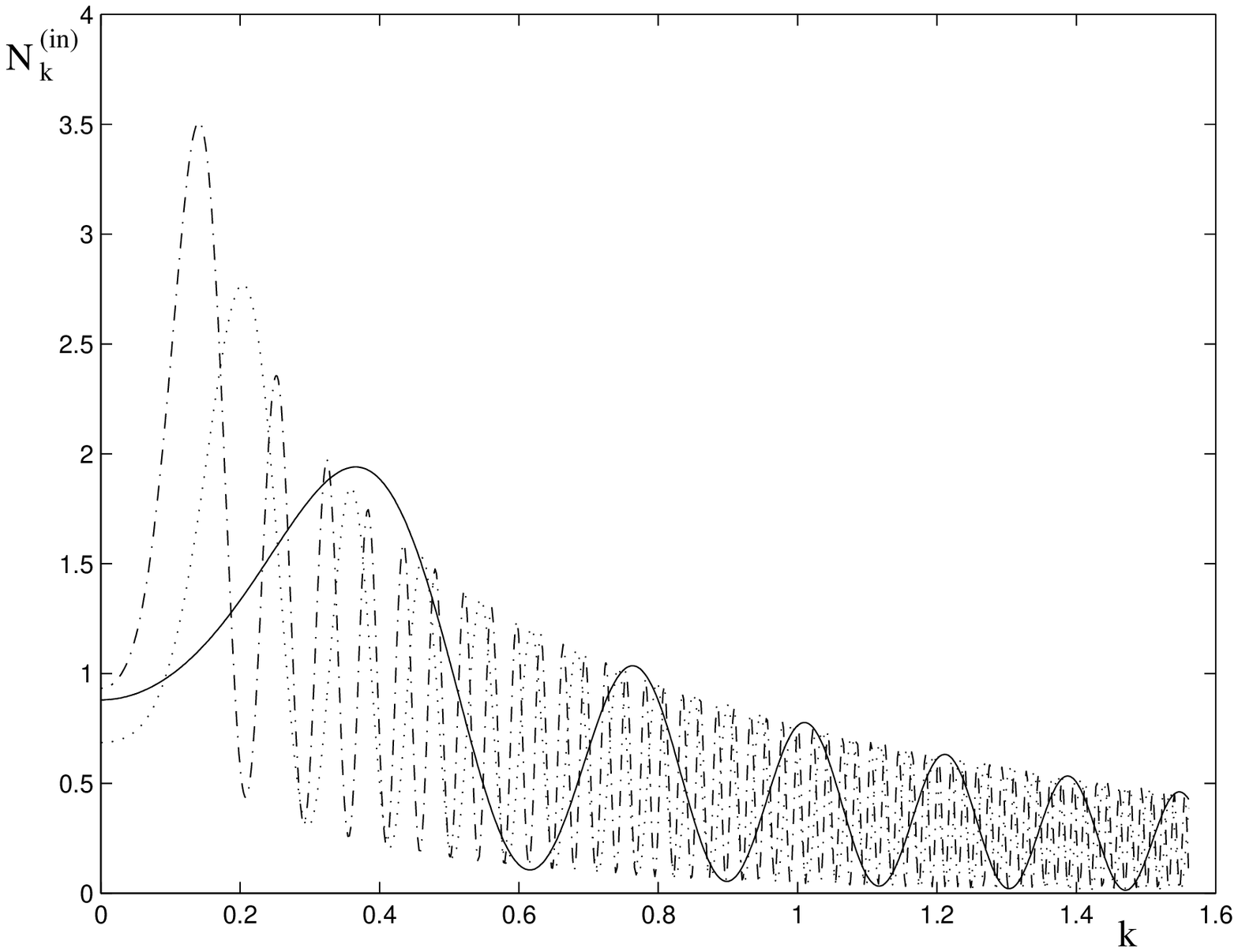,height=9cm,width=15cm}
\caption{\it Spectrum with respect to the initial vacuum, for $tm_{\rm
eq}=39.723$ (solid line), $199.006$ (dotted line) and $398.11$ (dashdot
line), for $\L/m_{\rm eq} = 20$, $\ell=0$ and $\rho_0=0.5$.}
\label{inspec}
\end{figure}

\begin{figure} 
\epsfig{file=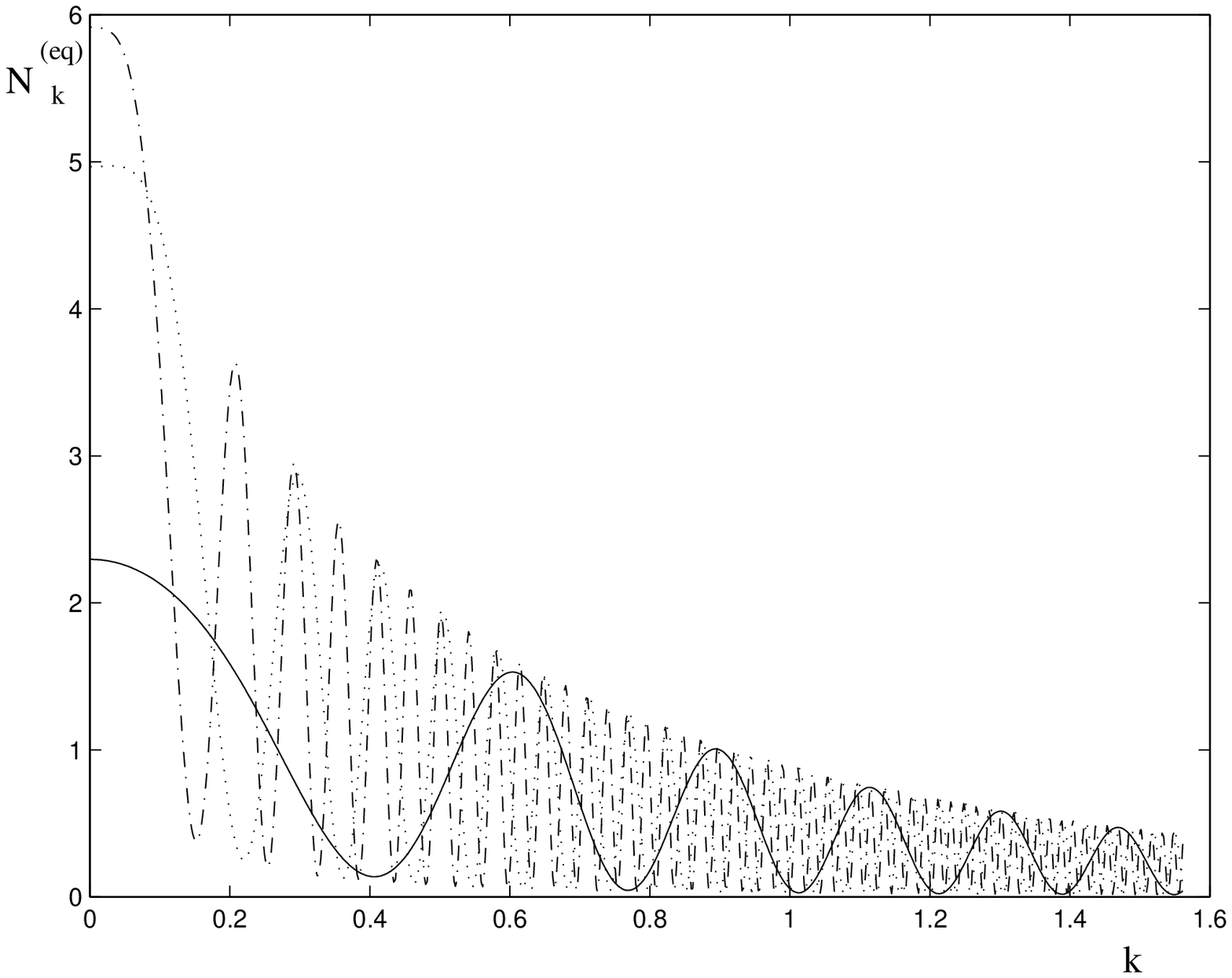,height=9cm,width=15cm}
\caption{\it Spectrum with respect to the true vacuum, for $tm_{\rm
eq}=39.723$ (solid line), $199.006$ (dotted line) and $398.11$ (dashdot
line), for $\L/m_{\rm eq} = 20$, $\ell=0$ and $\rho_0=0.5$.}
\label{eqspec}
\end{figure}

\begin{figure} 
\epsfig{file=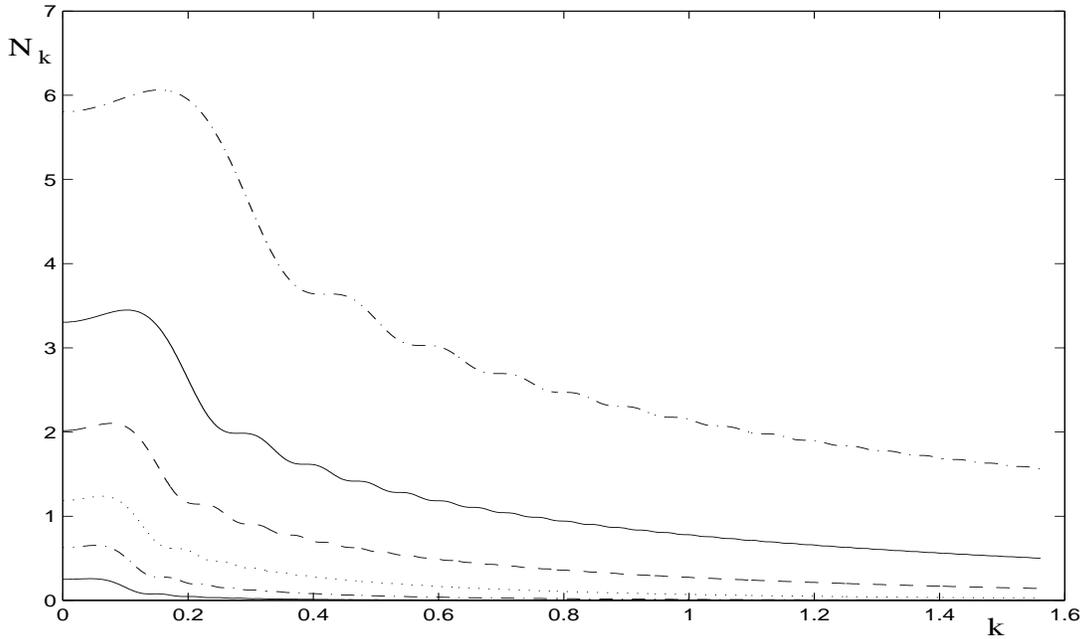,height=9cm,width=15cm}
\caption{\it Adiabatic spectrum for $tm_{\rm eq}=398.11$, $\L/m _{\rm
eq} = 20$ and $\ell=0$. The different curves correspond to different
initial values for the condensate: from top to bottom, $\rho_0=0.7$,
$0.6$, $0.5$, $0.4$, $0.3$ and $0.2$.}
\label{asy_sp}
\end{figure}

Once the evolution equations for the complex mode functions has been
solved, it is possible to compute the spectrum of the produced
particles. First, we should say that the notion of particle number is
ambiguous in a time dependent situation. Nevertheless, we may give a
suitable definition with respect to some particular pointer state. We
choose here two particular definitions, the same already used in the
study of the $\Phi^4$ model \cite{relax}, plus a third one. The first
choice corresponds to defining particles with respect to the initial
Fock vacuum state, the second with respect to the instantaneous
adiabatic vacuum state, and the third to the equilibrium vacuum (the
true vacuum of the theory). The corresponding expressions in terms of
the complex mode functions are:
\begin{equation}
\begin{split}
	&N^{\rm in}_k(t) = \frac14 \left[ \om_k |z_k (t)|^2 
		+ \frac{|\dot{z}_k(t)|^2}{\om_k} \right] - \frac12 \\
	&N^{\rm ad}_k(t) = \frac14 \left[ \om^{\rm ad}_k |z_k(t)|^2 + 
		\frac{|\dot{z}_k(t)|^2}{\om^{\rm ad}_k} \right] -\frac12
		\;,\quad \om^{\rm ad}_k = \sqrt{k^2+m(t)^2} \\
	&N^{\rm eq}_k(t) = \frac14 \left[ \om^{\rm eq}_k |z_k(t)|^2 + 
		\frac{|\dot{z}_k(t)|^2}{\om^{\rm eq}_k} \right] - \frac12
		\;,\quad \om^{\rm eq}_k = \sqrt{k^2+m_{\rm eq}^2} \\
\end{split}
\end{equation}
We report our numerical findings on these quantities in
figs. \ref{spec_1} - \ref{asy_sp}. Since the Lagrange multiplier tends
asymptotically to a constant value $m^2_{\infty}$, the condensate
$\rho(t)$ oscillates with frequency $m_{\infty}$ and the mode
functions $z_q(t)$ with frequency $\omega (q) =
\sqrt{q^2+m^2_{\infty}}$. This implies that particle spectra $N^{\rm
in}_k(t)$ and $N^{\rm eq}_k(t)$ are more and more strongly modulated
as time elapses, as figs. \ref{inspec} and \ref{eqspec} show; on the
contrary, $N^{\rm ad}_k(t)$ is a slowly varying function of the
momentum $k$ (cfr. figs. \ref{spec_1} - \ref{adnum0_6}), because the
oscillations of the mode functions are counterbalanced by the time
dependence of the adiabatic frequencies $\sqrt{k^2+m(t)^2}$. Finally,
fig. \ref{asy_sp} allows for a comparison of the spectra related to
different initial values of the condensate. 

Looking at the momentum distribution of the created particles at
different times, we see the formation of a growing peak corresponding
to soft modes. We can give an analytic, self-consistent description of
this behavior at large times through a perturbative approach, similar
to the one used in ref. \cite{relax}. We split the time--dependent
Lagrange multiplier in two parts, as in equation (\ref{asymass}) and
we treat the ``potential'' $p(t)/t$ perturbatively, as is done in
\cite{relax}. We find the following solution:
$$
z_q(t)=A_q e^{i\omega_qt} + B_q e^{-i\omega_qt} - \int_t^{\infty}
\frac{\sin \omega_q (t' - t)}{\om_q} \frac{p(t')}{t'} z_q(t') dt'
$$
which is equivalent, up to terms of order $O(1/t^2)$, to
\begin{equation}\label{pert_z}
\begin{split}
z_q(t) & =A_q \left[ 1 + \frac{A \sin \Psi (t)}{4im_{\infty}t} -
\frac{A}{8\omega_qt} \left( \frac{e ^{i \Psi (t)}}{\omega_q +
m_{\infty}} + \frac{e ^{-i \Psi (t)}}{\omega_q - m_{\infty}} \right)
\right] e ^{i \omega_q t} \\
& + B_q \left[ 1 - \frac{A \sin \Psi (t)}{4im_{\infty}t} -
\frac{A}{8\omega_qt} \left( \frac{e ^{i \Psi (t)}}{\omega_q -
m_{\infty}} + \frac{e ^{-i \Psi (t)}}{\omega_q + m_{\infty}} \right)
\right] e ^{- i \omega_q t} + O\left(\frac{1}{t^2}\right)
\end{split}
\end{equation}
with $\Psi (t) = 2 m_{\infty} t + \gamma_1 \log t +
\gamma_2$. The logarithmic dependence is due to the ``Coulomb form''
of the perturbative term $p(t)/t$ in the equations of motion.  The
expression (\ref{pert_z}) displays resonant denominators for
$\omega_q=m_{\infty}$, that is $q=0$.  The perturbative approach is
valid as long as the first order correction is small compared to
zeroth order. Such a condition is satisfied if
\begin{equation}
	\frac{{\cal A}}{4t\,\omega_q(\omega_q - m_{\infty})} < 1
\end{equation}
that implies $q^2>{\cal A}/4t$ for non relativistic modes. Thus the
position of the peak found before may be interpreted as the result of
a weak nonlinear resonance. The asymptotic behavior of the condensate
and the mode functions related to soft momenta must be obtained
through non-perturbative techniques, implementing a multitime scales
analysis and a dynamical resummation of sub-leading terms. A
self-consistent justification of the numerical result (\ref{pert1}),
along with the power law relaxation behavior for the expectation value
(with non-universal dynamical anomalous dimensions), are likely to be
obtained following the line of the analysis performed in \cite{relax}
for the $\phi^4$ model in $3+1$ dimensions.

From the numerical study of the complete spectrum history, we conclude
that no exponentially growing (parametric or spinodal) instabilities
are present in the case at hand, as apparent from Fig.s \ref{spec_1} -
\ref{adnum0_6}, which show the spectrum of produced particles with
respect to the adiabatic vacuum state.

\section{Outlook}
\label{ol}
The natural continuation of this paper is the detailed numerical study
of the evolution, in order to give a precise picture of the process of
mean field dissipation via particle production in the framework of
this constrained, asymptotically free model. It should be possible to
determine precisely the power laws that characterize the asymptotic
evolution of relevant variables, like the condensate, the Lagrange
multiplier and the number of created particles. After this, one should
be able to decide whether, at zero angular momentum, the damping leads
to the complete dissipation of the energy stored in the condensate or
the system evolves towards a limit cycle with an asymptotic amplitude
different form $0$. Also a comparison with the linear model in $1+1$
dimension might be useful to understand the peculiarities of the
dynamics in a constrained model.

Moreover, it would be very interesting to study the dependence of the
evolution on the value of $\ell$, the angular momentum of the field in
the internal space. As the preliminary results presented in this paper
show (see figure \ref{lneq0_3}), the asymptotic state is far from the
state of minimum energy compatible with the given value of
$\ell$. Remarkably, the adiabatic spectrum of produced particles in
case of $\ell \neq 0$ is broader than that one corresponding to
$\ell=0$, suggesting a stronger coupling with hard modes.

Most important, the effects of the inclusion of $O(1/N)$ corrections
and the evolution of non unigorm condensates should be analyzed also
in the framework of the nonlinear $\sigma$ model in $1+1$ dimension.

\subsection*{Acknowledgements}
E.M. thanks Italian MURST and INFN for financial support.
\vskip 0.5 truecm


\vskip 0.5 truecm

\vskip 0.5 truecm

\vskip 0.5 truecm

\vskip 0.5 truecm

\vskip 0.5 truecm

\vskip 0.5 truecm




\vskip 0.5 truecm


\vskip 0.5 truecm

\vskip 0.5 truecm

\vskip 0.5 truecm

\vskip 0.5 truecm

\vskip 0.5 truecm

\end{document}